\documentclass[12pt]{article}
\usepackage{amsthm,amsmath,amsfonts,amssymb,cases}
\usepackage{comment}
\usepackage{enumerate}
\usepackage{hyperref}
\usepackage{cleveref}
 \usepackage{tikz}
\usepackage{pgfplots}
\usepackage[shortlabels]{enumitem}
\usepackage{float}

\usepackage[style=alphabetic,giveninits=true,maxnames=4, backend=biber,doi=false,url=false,isbn=false,eprint=false]{biblatex}
\newcommand{\R}{{\mathord{\mathbb R}}}
\newcommand{\Z}{{\mathord{\mathbb Z}}}
\newcommand{\N}{{\mathord{\mathbb N}}}
\newcommand{\C}{{\mathord{\mathbb C}}}

\newcommand{\E}{{\mathord{\mathbb E}}}

\newcommand{\HH}{\mathcal{H}}

\newcommand{\muun}[1]{\begin{align} #1 \end{align}}

\newcommand{\SP}[1]{\left\langle #1  \right\rangle }
\renewcommand{\epsilon}{\varepsilon}

\newcommand{\vertiii}[1]{{\left\vert\kern-0.25ex\left\vert\kern-0.25ex\left\vert #1
\right\vert\kern-0.25ex\right\vert\kern-0.25ex\right\vert}}

\newcommand{\ben}{\begin{displaymath}}
\newcommand{\een}{\end{displaymath}}
\newcommand{\beqn}{\begin{equation}}
\newcommand{\eeqn}{\end{equation}}
\newcommand{\beqna}{\begin{eqnarray*}}
\newcommand{\eeqna}{\end{eqnarray*}}

\newcommand{\nn}{\nonumber}

\newcommand{\inn}[1]{\left\langle {#1} \right\rangle }

\newtheorem{lemma}{Lemma}
\newtheorem{theorem}[lemma]{Theorem}
\newtheorem{remark}[lemma]{Remark}
\newtheorem{proposition}[lemma]{Proposition}
\newtheorem{corollary}[lemma]{Corollary}
\newtheorem{definition}[lemma]{Definition}
\newtheorem{hyp}{Hypothesis}

\numberwithin{equation}{section}
\numberwithin{lemma}{section}

\makeatletter
\newsavebox{\@brx}
\newcommand{\llangle}[1][]{\savebox{\@brx}{\(\m@th{#1\langle}\)}%
  \mathopen{\copy\@brx\kern-0.5\wd\@brx\usebox{\@brx}}}
\newcommand{\rrangle}[1][]{\savebox{\@brx}{\(\m@th{#1\rangle}\)}%
  \mathclose{\copy\@brx\kern-0.5\wd\@brx\usebox{\@brx}}}
\makeatother

\begin{document}
\title{Approximating     the density of states   for  Poisson distributed random Schr\"odinger operators}
\author{\vspace{5pt} David Hasler$^1$\footnote{
E-mail: david.hasler@uni-jena.de } \quad  Jannis Koberstein $^1$\footnote{E-mail: jannis.koberstein@eah-jena.de.} \\
\vspace{-4pt} \small{$1.$ Department of Mathematics,
Friedrich Schiller  University  Jena} \\ \small{Jena, Germany } }
\date{}
\maketitle

\begin{abstract}
We consider a   Schr\"odinger operator with random potential distributed according
to a Poisson process. We show that expectations of matrix elements of the resolvent as well as
the density of states can be approximated to arbitrary precision
in powers of the coupling constant.
The expansion coefficients  are  given in terms
of    expectations  obtained by    Neumann expanding the potential around the free Laplacian.
One  can control  these expansion coefficients in the
infinite volume limit. We show  that for this
limit the boundary value as the  spectral
parameter approaches the real axis exists as well. Our results are valid for arbitrary strength of the disorder parameter,
including  the small disorder regime.
\end{abstract}


\section{Introduction}


Random Schr\"odinger operators are a class of  mathematical models  describing transport properties in solids, which are subject to
some form of disorder.
Starting  with the seminal paper of Anderson  \cite{Anderson.1958} in 1958 such models have intensively been investigated.
Various physical as well as mathematical properties  have been shown for this model.
In particular various localization properties
have been established mathematically  rigorous, starting with \cite{FrohlichSpencer.1983} see also \cite{AizenmanMolchanov.1993}. Specifically, for large disorder existence of   dense pure point spectrum with exponentially localized eigenfunctions has been proven. In this paper we focus on the small disorder regime. In this regime much less is known.
One  conjectures that this regime exhibits extended states, i.e., absolutely continuous spectrum. A few results have been obtained supporting
this conjecture, cf. \cite{AcostaKlein.1992,Klein.1998,FroeseHaslerSpitzer.2006,AizenmanSimsWarzel.2006,FroeseHaslerSpitzer.2007,AizenmanWarzel.2013} for results on the Bethe lattice. Nevertheless, the common belief is that there is a  phase transition of some form between the regime of large disorder and
small disorder.


The density of states has been studied intensively for such operators.
The boundedness of the density of states follows from a so called optimal  Wegner estimate.
A Wegner estimate \cite{Wegner.1981} gives an upper bound on the number of expected eigenvalues in an interval.
It is optimal if the bound is proportional to the volume and the interval length.
Wegner estimates have been obtained for various models, cf. \cite{KirschMartinelli.1982,CombesHislopKlopp.2003,CombesHislopKlopp.2007,KirschMetzger.2007,ErdosHasler.2012}. In particular,
for Poisson distributed random Potential a Wegner estimate has been obtained
which is  optimal in the interval size but not in the volume \cite{CombesHislop.1994}. Note that it is possible
to turn Wegner estimates which are not optimal in the volume into an optimal in the volume estimate
 in exchange for optimality in the interval length, cf. \cite{HaslerLuckett.2013}.
For typical Wegner estimates,  the proportionality constant is of the magnitude of the
inverse  disorder strength. This gives a good estimate  for large disorder.
However for small disorder, such  estimates  do not seem  optimal.
For this reason we follow  a different approach.

 In this paper, we study a Laplacian in $\R^d$ with a Poisson distributed random potential.
 Such  random potentials are used to  model amorphous materials like glass or rubber.
We consider
expansions of the expectation value of the interacting resolvent in terms of the free resolvent.
In particular, we show that the expectation of the interacting resolvent can be approximated  in terms of expectations involving the free resolvent with
arbitrary precision.  We then show
that  the density of states can be approximated to arbitrary precision using this expansion as well.
The estimates are uniform in the disorder strength in a bounded interval around zero.
 Finally, we
show that the  expansion coefficients  have infinite volume limits
which are finite as the spectral parameter approaches the real axis within the spectrum
of the free Laplacian.

The present paper complements the work \cite{haskob.2022}. While the main result, Theorem \ref{thm:duhamelmainexp} stands completely on its own,  Corollary \ref{thm:duhamelmainexpdens} uses results from \cite{haskob.2022}.
To obtain our result, we expand the resolvent in a Neumann Series by treating the random
potential as a perturbation of the free Laplacian. Then we calculate the expectation
and show that these expansion coefficients are finite in an infinite volume limit.
The techniques which we use are inspired by \cite{ErdosSalmhoferYau.2007a,ErdosSalmhoferYau.2007b,ErdosSalmhoferYau.2008a}. A related expansion
method of the resolvent has been used in  \cite{MagnenPoirotRivasseau.1996,MagnenPoirotRivasseau.1998,Poirot.1999}.
A typical challenge in such an analysis is to obtain  a suitable   estimate on the  combinatorics of the expansion terms.
In this paper we   focus directly on  the error term.
The error estimate of our  main result shows that this expansion technique
gives the correct answer to arbitrary order.


This manuscript is structured as follows. In the next section we introduce the model and state the main results.
The proofs are given in  Section   \ref{sec:proofofmainthms}. In the appendix we collect
a few results, which we need for the proofs.

\section{Model and Statement of Main Results}
\label{Model}

In this section we  introduce  the model and state the main results on the expectation
of the resolvent.  We note that the definition of the model follows  the one given in \cite{ErdosSalmhoferYau.2008a} closely.
We consider   a finite box of size $L  > 0$ and define the  box
 $\Lambda_L := \left[ - \frac{1}{2} L , \frac{1}{2} L \right)^d \subset \R^d$, where $d \in \N$ denotes the spacial dimension.
Let $\inn{ \cdot , \cdot }_{L^2(\Lambda_L)}$ and $\| \cdot \|_{L^2(\Lambda_L)}$ denote the canonical scalar product and the norm of the Hilbert space  $L^2(\Lambda_L)$. If it is clear from the
context what the inner product or  the norm is,   we shall occasionally drop the subscript $L^2(\Lambda_L)$.
To introduce the periodic Laplacian operator on this space it is convenient to
work in terms of Fourier series.  For this, let
$\Lambda_L^*  = ( \frac{1}{L} \Z  )^d=\{ (k_1 ,  \ldots ,  k_d) : \forall j=1, \ldots , d,  \exists m_j \in \Z,: k_j =\frac{m_j}{L} \}$ denote
the so-called dual lattice. We introduce the notation
\begin{equation}
\int_{\Lambda_L^*} f(p) dp  := \frac{1}{|\Lambda_L |} \sum_{p \in \Lambda_L^*} f(p) .
\end{equation}
The sum  $\int_{\Lambda_L^*}f(p)dp$ can be interpreted as a Riemann-sum, which  converges to the integral $\int f(p) dp $
as   $L \to \infty$, provided $f$ has  sufficient  decay at infinity and sufficient regularity.
For a subset $S \subset \Lambda_L^*$ we set  $\int_{S,*} f(p)  dp  := | \Lambda_L|^{-1} \sum_{p \in S} f(p)$.
Furthermore, we introduce the notation for the canonical norm in the $\ell^q$-spaces.  For  $f \in \ell^q(\Lambda_L^*)$
with $q  \in [1,  \infty]$,
we write
  \begin{align*} \| f \|_{*, q} = \left( \int_{\Lambda_L^*} |f(k)|dk \right)^{1/q}  \ ( 1 \leq q < \infty)   , \quad   \| f \|_{*, \infty } = \sup_{k \in \Lambda_L^*} |f(k)| .
  \end{align*}
{ For compactness of notation we shall denote $\| \cdot \|_{*,2}$ simply by $\| \cdot \|_*$.  }

For any $f \in L^1(\Lambda_L)$ we define the Fourier series
$$
\hat{f}(p) = \int_{\Lambda_L} e^{ - 2 \pi i p \cdot x}  f(x) dx \text{ for } p \in \Lambda_L^*  .
$$
  Further, for $g    \in \ell^1(\Lambda_L^*)$ we define the inverse
$$
\check{g}(x) = \int_{\Lambda^*_L} e^{  2 \pi i  p \cdot x } g(p) dp  = \frac{1}{|\Lambda_L |} \sum_{p \in \Lambda_L^*} e^{  2 \pi i  p \cdot x } g(p) \text{ for }  x \in \Lambda_L.
$$
Note that
$\check{(\cdot)}$ extends uniquely to a continuous linear map  from $ \ell^2(\Lambda_L^*)$ to $ L^2(\Lambda_L)$.
We shall denote this extension again by the same symbol. This extension maps $\ell^2(\Lambda_L^*)$ unitarily to $L^2(\Lambda_L)$
and it is the inverse of $\hat{(\cdot)}|_{L^2(\Lambda_L)}$, see for example \cite[Theorem 8.20]{folland}.  In physics literature,  one might be familiar with the notation where  $L^2(\Lambda_L)$ represents the position space,  while $\ell^2(\Lambda_L^*)$ would be called momentum space,  or Fourier space.
For $ p \in \Lambda_L^*$  and $x \in \Lambda_L$, we define
\begin{align} \label{eq:deofonb}
 \varphi_p(x) ={|  \Lambda_L|}^{-1/2} e^{ i 2 \pi p \cdot x } .
\end{align}
Observe that
$
\{ \varphi_p :p \in \Lambda_L^* \}
$
is an orthonormal Basis (ONB) of $L^2(\Lambda_L)$ \cite[Theorem 8.20]{folland} and that
$\hat{f}(p) = | \Lambda_L|^{1/2}  \inn{  \varphi_p , f}_{L^2(\Lambda_L)} $.

When restricting the Laplacian  to a finite box, we choose periodic boundary conditions, which will be technically convenient when working in momentum space.
Therefore,  we are going to introduce the Laplacian with periodic boundary conditions on $L^2(\Lambda_L)$ by means of Fourier series.
In this paper, we shall adapt the standard convention that for a vector $a =(a_1,...,a_d)$ we use the notation $a^2 := |a|^2$ where $|a| :=  \sqrt{ \sum_{j=1}^d |a_j|^2}$.
In that sense,  we define the energy function
\begin{equation}
\label{energiyfunction}
\nu : \R^d  \to \R_+,  \quad p \mapsto \nu(p) := \frac{1}{2} p^2,
\end{equation}
which we can now use to
 define $-\Delta_L$ as the linear operator with domain
\begin{align*}
D(-\Delta_L) = \{ f \in L^2(\Lambda_L) : \nu \hat{f} \in \ell^2(\Lambda_L^*) \},
\end{align*}
and the mapping rule
\begin{equation} \label{eq:deoflaplac0}
 -\Delta_L : D(-\Delta_L)  \to L^2(\Lambda_L), \quad f \mapsto -\Delta_L f   =  (2 \pi )^2 ( \nu  \hat{f})^\vee .
\end{equation}
Observe that  $-\Delta_L$ is self-adjoint, since it is  unitary equivalent to  a multiplication operator by a real valued function,  and that we have the identity
\begin{equation} \label{eq:deoflaplac}
\left[ - \frac{1}{2 (2 \pi)^2}   \Delta_L  f \right]^\wedge(p)  = \nu (p) \hat{f}(p)  .
\end{equation}
By
\begin{equation} \label{defofmainH}
 H_{L} := H_{\lambda,L} := - \frac{\hbar^2}{ 2 m } \Delta_L + \lambda V_{L}
\end{equation}
we denote a random Schr\"odinger operator acting on $L^2(\Lambda_L)$ with   a random potential
$
V_{L} = V_{L,\omega}(x)
$, defined below,  and a  coupling constant $\lambda \geq 0$.   As in  \cite{ErdosSalmhoferYau.2008a}  we choose units for the mass $m$ and Planck's constant  so that $\frac{\hbar^2}{2m} = [2 (2\pi)^2]^{-1}$.
For a function $h : \R^d \to \C$ we denote by $h_\#$ the $L$--periodic extension of $h|_{\Lambda_L}$ to $\R^d$.
The potential is given by
\begin{align} \label{defofpot101}
V_{L,\omega}(x) := \int_{\Lambda_L} B_\#(x - y) d \mu_{L,\omega}(y),
\end{align}
where $B$,  having the physical interpretation as   a single site potential profile,   is  assumed  to be a  real valued Schwartz function on $\R^d$.
Moreover, we assume that either $B$  has   compact support
or that  $B$  is  symmetric with respect to the reflections of the coordinate axis,  i.e.  for  $j=1,...,d$,
\muun{ \label{SC}
\mathfrak{S}_j:  (x_1,...,x_j,...,x_d) &\mapsto (x_1,...,-x_j,...,x_d)
\\
B \circ \mathfrak{S}_j &= B. \nn
}
\begin{remark}
Each of the two conditions is mathematically convenient in the sense that they ensure sufficiently fast decay of the Fourier transform.  Further the reflection symmetry condition is satisfied by rationally invariant potentials, which occur naturally.
\end{remark}
Furthermore,  let $\mu_{L,\omega}$ be a Poisson point measure  on $\Lambda_L$ with homogeneous unit density and
with independent indentically distributed (i.i.d.) random masses.
More precisely, for almost all events $\omega$, it consists of $M$ points $\{ y_{L,\gamma}(\omega) \in \Lambda_L : \gamma=1,2,...,M\}$, where $M = M(\omega)$ is a Poisson variable with expectation $|\Lambda_L|$, and $\{ y_{L,\gamma}(\omega) \}$ are i.i.d. random variables uniformly distributed on $\Lambda_L$.   Both are independent of  the random i.i.d. weights  $\{ v_\gamma : \gamma=1,2,... \}$ and
the random measure is given by
$$
\mu_{L,\omega} = \sum_{\gamma=1}^{M(\omega)} v_\gamma(\omega) \delta_{y_{L,\gamma}(\omega)} ,
$$
where $\delta_y$ denotes the Dirac mass at  the point $y$.
Note that the case where $M(\omega )=0$ corresponds to a vanishing potential.
Note that for convenience the notation  will not reflect the   dependence on the specific  choice of the common  distribution of the weights  $\{ v_\gamma \}$.   We assume that the moments
\muun {\label{moments}
m_k :=  {\bf E} v_\gamma^k
}
 satisfy
\begin{equation} \label{2.4}
 m_{k } < \infty     , \quad  \text{ for all }  k \in \N.
\end{equation}
For some of the results we will use that the first moment $m_1 = {\bf E} v_\gamma$ vanishes.
Observe that we can write   \eqref{defofpot101} as
\begin{equation}\label{bumpnotationres}
 V_{L,\omega}(x) =  \sum_{\gamma=1}^M V_{L,\gamma}(x)   \quad \text{ with } \quad V_{L,\gamma}(x) := v_\gamma B_\#(x-y_{L,\gamma}) .
\end{equation}
The expectation with respect to the joint measure of $\{ M , y_{L,\gamma}, v_\gamma\}$ is denoted by ${\bf E}_L$.
Sometimes we will use the notation
\begin{equation}\label{esy:3.24}
{\bf E}_L = {\bf E}_M {\bf E}_{y_L}^{\otimes M } {\bf E}_v^{\otimes M }
\end{equation}
referring to the expectation of $M$, $\{ y_{L,\gamma} \}$ and $\{ v_\gamma \}$ separately.
In particular, ${\bf E}_y^{\otimes M }$ stands for the normalized integral
\begin{equation}\label{esy:3.240}
\frac{1}{|\Lambda_L|^M} \int_{(\Lambda_L)^M} dy_1 \cdots dy_M .
 \end{equation}
Since the potential is almost surely bounded it follows  by standard perturbation theorems, e.g. the Kato-Rellich theorem \cite{SR2},  that   the operator \eqref{defofmainH}  is almost
surely self-adjoint  for all $\lambda \geq 0$.

The main object of our interest is the  expectation  of  matrix values of the resolvent
\muun{
\label{GFct}
{\bf E}_L  \langle \psi_1, ( H_{ L}  - z)^{-1}   \psi_2 \rangle \quad \text{ for } \psi_1 , \psi_2 \in L^2 (\Lambda_L)
}
as the spectral parameter approaches the real axis.
Note that  \eqref{GFct}  is the expectation of the interacting resolvent, which is difficult to study.  To analyze  \eqref{GFct}  we will  use a Neumann expansion to express the interacting  resolvent as the sum of products of powers of the potential and the resolvent of the periodic Laplacian, $\Delta_L$.
For notational compactness   we  shall denote   the resolvent of $-\Delta_L$  by
$$
R_L(z ) :=  \left(- \frac{\hbar^2 }{2 m} \Delta_L - z \right)^{-1} ,
$$
where $z \in \C \setminus [0,\infty)$.
Note that    we use a   notation for the   resolvent of the free Laplacian, which   one might expect  for the interacting  one.

 For  $z \in \C \setminus [0,\infty)$ and $\psi_1, \psi_2 \in L^2(\R^d)$ we define
 \begin{align} \label{defoffinT}
  T_{n,L}[z;\psi_1,\psi_2]
& :=  {\bf E}_L \inn{ \psi_{1,\#},    R_L(z) [  V_L  R_L(z)  ]^n \psi_{2,\#} }  .
\end{align}
By definition the potential is almost surely bounded. Thus  \eqref{defoffinT} is well defined almost surely.
To state the results we  will  define the infinite volume Fourier transform and inverse Fourier transform for $f \in L^1 (\R^d) $
\begin{align*}
\hat{f}(k) = \int_{\R^d} f(x) e^{ - 2 \pi i k \cdot  x } dx  , \quad k \in \R^d ,
\end{align*}
and
\begin{align*}
\check{f}(x) =\int_{\R^d}   f(k) e^{  2 \pi i k \cdot  x } dk, \quad x \in \R^d ,
\end{align*}
respectively.

   Next we state a result that shows that the    matrix elements of  the resolvent of the Hamiltonian can be calculated to arbitrary precision in terms of the expansion coefficients
  $T_{n,L}$ defined in  \eqref{defoffinT}.
The error estimate which we obtain is  uniform in the volume $\Lambda_L$.
To obtain the result we use a Du Hamel expansion and a combinatorical estimate.
To formulate the main theorems we introduce
the following hypothesis and the subsequent definition.
\begin{hyp} \label{H4} The random potential $v$ satisfies that there exists a constant $C$ such that
$$
m_n := {\bf E} (  |v|^n )  \leq C
$$
for all $n \in \N $.
\end{hyp}
In addition we  will work   with a  regularization function, whose properties are as  follows.
\begin{definition} \label{H3}
Let $\chi : \R \to [0,1]$ is  a $C^\infty_c(\R)$ be a function such $\chi =1$ on $[-1,1]$ and $\chi(t) = 0$ for $|t| \geq 2$.  Let $\chi_a(t) := \chi(at)$ for all $a,t \in \R$.
\end{definition}
A proof of the existence of such a regularization function can be found in \cite[Theorem 1.4.1]{Hoermander.1989} for example.
We can now state the first  main   result of this paper. 
\begin{theorem} \label{thm:duhamelmainexp}
Suppose Hypothesis {\bf \ref{H4}} holds.  Let     $\eta > 0$,  $\epsilon > 0$,  and $\lambda_0 > 0$.
Then
there exists an   $N= N(\epsilon,\eta, \lambda_0)  \in \N$   such that for all  $\psi_1,\psi_2 \in L^2(\R^3)$, $E  \in \R$,
$\lambda \in [-\lambda_0, \lambda_0]  $, and $L \geq 1$
\begin{align}  \label{firsteqmaincont007}
\left| {\bf E}_L \inn{ \psi_{1,\#} , (H_\lambda - (E  +    i \eta ) )^{-1}      \psi_{2,\#} }  -   \sum_{n=0}^{N}  \lambda^n S_{n,L}[ E +  i   \eta; \chi; \psi_1, \psi_2 ; \epsilon ]    \right| \leq \epsilon  \| \psi_{1,\#} \|_* \| \psi_{2,\#} \|_*,
\end{align}
where we defined
\begin{align} \label{SnL}
S_{n,L}[ E  +  i \eta  ; \chi ; \psi_1, \psi_2; \epsilon  ]  := (-1)^n  \int_\R \widehat{\chi}_{a(\eta , \epsilon)}(\alpha)  T_{n,L}[ E +  i \eta  +  2\pi \alpha  ; \psi_1, \psi_2 ]   d \alpha  ,
\end{align}
and   \begin{align}  a := a(\eta,\epsilon)  := \frac{  \eta}{ | \ln( \frac{\eta \epsilon}{2})| + 1 } \label{defofsca} .  \end{align}
\end{theorem}

The proof of \ref{thm:duhamelmainexp} will be given in Section  \ref{sec:proofofmainthms}.

\begin{remark} {\rm
We note that the typical  form of the so called  Wegner estimate gives an upper bound of the order $|\lambda|^{-1}$.  The trivial estimate \ref{ln-1} gives a bound $\eta^{-1}$.
Thus by interpolation we find that the resolvent in the above expression is bounded by $\lambda^{-1+c} \eta^{-c}$ for any $c \in [0,1]$.
Theorem  \ref{thm:duhamelmainexp}  allows to  choose $\epsilon$ arbitrary small, e.g., much smaller than $\lambda^{-1+c} \eta^{-c}$.
 Thus Theorem  \ref{thm:duhamelmainexp} allows to  determine the resolvent to arbitrary good precision.}
\end{remark}

 As a corollary of Theroem \ref{thm:duhamelmainexp} we can approximate the density of states up to
arbitrary precision. To formulate the corollary we introduce the notation
\begin{align} \label{fEeta}
f_{E,\eta}(x) =  \frac{\eta}{(x-E)^2 + \eta^2} .
\end{align}
For this we assume  $d \leq 3$, which  is  necessary  to describe the density of states in terms of the real part of the
resolvent, by means of \eqref{fEeta}, and to ensure at the same time that the   trace in  \eqref{firsteqmaincont007D}
is  well defined.  Let ${\rm Tr}_L$ denote the trace with respect to $L^2 (\Lambda_L)$.

\begin{corollary} \label{thm:duhamelmainexpdens} Let $d \leq 3$. Suppose Hypothesis {\bf \ref{H4}} holds.
Assume that the support of the distribution of $v_\alpha$ is contained in $(0,\infty)$.
 Let   $\eta > 0$,
 $\epsilon > 0$, $E > 0$,  and $\lambda_0 > 0$.
Then there exists    $N = N(\epsilon,\eta, \lambda_0)   \in \N$  and a  number $\kappa = \kappa(\epsilon,\eta,\lambda_0) > 0$
  such that for all $\lambda \in [0, \lambda_0]  $, $E \in \R$,  and $L \geq 1$
\begin{align}  \label{firsteqmaincont007D}
\left| \frac{1}{|\Lambda_L|} {\bf E}_L  {\rm Tr}_L f_{E,\eta} ( H_\lambda)  -   \sum_{n=0}^{N} \int_{|q| \leq \kappa,*} \lambda^n  {\rm Im} S_{n,L}[ E + i   \eta; \chi; \varphi_{q}, \varphi_q ; \epsilon ] dq    \right| \leq \epsilon
\end{align}
with coefficients defined in \eqref{SnL} and \eqref{defofsca}.
\end{corollary}
The proof of Corollary \ref{thm:duhamelmainexpdens} will be given in Section  \ref{sec:proofofmainthms}.

\begin{remark} {\rm  Although the estimate of Corollary \ref{thm:duhamelmainexpdens}
allows us to approximate the density of states to arbitrary precision in terms of expectations of products of the free resolvent,
we do not have  a priori   control on the large $N$ behavior of the expression
\begin{align}\label{sumoverS}
\sum_{n=0}^{N} \int_{|q| \leq \kappa,*} \lambda^n {\rm Im}  S_{n,L}[ E + i   \eta; \chi; \varphi_{q}, \varphi_q ; \epsilon ] dq
\end{align}  uniformly in   $\eta \in (0,\eta_0]$ for some $\eta_0 > 0$.
It would be interesting to  establish control of the  growth  of \eqref{sumoverS} directly or its infinite volume limit,  cf. \cite{ErdosSalmhoferYau.2007a,  ErdosSalmhoferYau.2007b, ErdosSalmhoferYau.2008a} for an  analysis in this direction.
Nevertheless, the estimate  \eqref{firsteqmaincont007D} shows that the expansion \eqref{sumoverS} converges.  In fact, a
  Wegner estimate which is optimal in volume as well as interval length
for the Poisson distributed random  Potential,  as considered in this paper,
would  imply the convergence of the infinite volume limit of \eqref{sumoverS},  by means of \eqref{firsteqmaincont007D}.
}
\end{remark}

Finally,  we  analyze the expansion coefficients   \eqref{SnL} more closely using results from \cite{haskob.2022}.
First, we will show   that their infinite volume limit exists and, second, we show
that the infinite volume  limit  is  in fact finite as the spectral parameter approaches the real axis.
Before we address this,  we need to make sure that the limit $L \to \infty$ indeed exists.
To this end we use results from  \cite{haskob.2022}.
The next theorem shows that
for each expansion coefficient  the infinite volume
limit   exists and  gives an upper bound dependent on  the spectral parameter for each expansion coefficient .

\begin{theorem}[\cite{haskob.2022}]  \label{exinflim} Let  $n \in \N$.
For all  $z \in \C \setminus [0,\infty)$, and   $\psi_1 ,\psi_2 \in L^2(\R^d)$  the   limit
$ \lim_{L \to \infty} T_{n,L}[z; \psi_1, \psi_2] $ exists in $\C$. There exists a  constant $K_n$ such that  the inequality
\begin{align} \label{boundonTs}
| T_{n,L}[z;\psi_1,\psi_2 ] | \leq \frac{ K_n   \| \psi_1 \| \| \psi_2 \| }{ {\rm dist}(z,[0,\infty))^{n+1} }
\end{align}
holds for all $L \geq 1$, $z \in \C \setminus [0,\infty)$, and  $\psi_1, \psi_2 \in L^2(\R^d)$.
\end{theorem}

Due to Theorem    \ref{exinflim}, we can define the infinite volume
 limit of the expansion coefficients
\begin{equation} \label{eq:defoflimL}
T_{n,\infty}[z ; \psi_1,\psi_2] := \lim_{L \to \infty} T_{n,L}[z; \psi_1, \psi_2]  ,
\end{equation}
for any $\psi_1,\psi_2 \in L^2(\R^d)$  and  $z \in \C \setminus [0,\infty)$. Theorem \ref{exinflim} can also be used to calculate the $L \to \infty$ limit of $S_{n,L}$.

\begin{proposition} \label{thm:duhamelmainexp3}
Suppose Hypothesis {\bf \ref{H4}} holds.  Let $E > 0$ and $\eta > 0$. Then for  \eqref{SnL} with  \eqref{defofsca}
we find
\begin{align} \label{inflimS}
  &\lim_{L \to \infty} S_{n,L}[E + i \eta; \chi ; \psi_1,\psi_2;\epsilon]   =  \int_{\R}    \widehat{ \chi}_a(\alpha)  T_{n,\infty}[E +2 \pi \alpha +  i \eta ;  \psi_1 , \psi_2 ]    d \alpha.
\end{align}
\end{proposition}
\begin{proof}
Using  the estimate in   \eqref{boundonTs}  of      Theorem  \ref{exinflim}  we obtain an $L^1$-bound for the integrand on the LHS found in \eqref{SnL}.  Now the statement follows from the dominated convergence  theorem.
\end{proof}

 To state the   next  result  about the
expansion coefficients of the resolvent  \eqref{eq:defoflimL},
  we need some regularity assumptions for the profile function and the wave functions with respect to which
  we calculate the matrix element of the resolvent.
  These regularity assumptions are
  formulated in terms of so called analytic dilations, cf. \cite{CombesThomas.1973,SR4}.  Let us first introduce the group of dilations.

 \begin{definition}
 \label{defutheta}
The group of unitary operators $u(\theta)$ on $L^2(\R^d)$ given by
\muun{
\label{equtheta}
(u(\theta) \psi)(x) = e^{ \frac{d \theta}{2} } \psi(e^\theta x ) \quad \text{for} \ \theta \in \R ,
}
 is called the {\bf group of dilation operators on} $\R^d$.
 \end{definition}

 It is straightforward to verify that $\R \mapsto \mathcal{B}(L^2(\R^d))$, $\theta \mapsto u(\theta)$ is indeed a strongly continuous one parameter group of  unitary operators, cf. \cite[Page 265]{SR1}.
 Observe that from the definition of the Fourier transform and the dilation operator it is easy to see that  for any $\psi \in L^2(\R^d)$ and all $\theta \in \R$
 \begin{equation} \label{eq:dilfour}
 (u(\theta) \psi)^{\wedge} = u(-\theta) \hat{\psi}.
 \end{equation}
 Analogously to \cite{haskob.2022}  we  assume  the following hypotheses. First we state a  hypothesis about the profile function needed for Theorem  \ref{thm:boundaryvelueexpcoeff}, below.

\begin{hyp} \label{H1} There exists $\vartheta_B > 0$ such that the Fourier transform, $\hat{B}$, of the Schwartz function $B$ satisfies that for $p=1 $ and $p=\infty$ the  function $\R \to L^p(\R^d),  \theta \mapsto  \widehat{B}_{\theta} := (u(\theta) B)^\wedge$  has  an extension
to an analytic function
$D_{\vartheta_B} := \{ z \in \C : |z| < \vartheta_B\}\to L^p(\R^d).$
\end{hyp}
Let us now state the hypothesis about the wave function with respect to which we shall calculate matrix elements of the resolvent  in Theorem \ref{thm:boundaryvelueexpcoeff}  below.

\begin{hyp} \label{H2} For the vector $\psi \in L^2(\R^d)$ and there exists $\vartheta_\psi > 0$ such that the transform $\hat{\psi}$ satisfies that the  function  $\R \to L^p(\R^d),    \theta \mapsto \widehat{\psi}_\theta := \widehat{\psi}(e^\theta \cdot ) $ has  an extension
to an analytic function $D_{\vartheta_\psi}  := \{ z \in \C : |z| < \vartheta_\psi\} \to L^2(\R^d)$.
\end{hyp}

In \cite{haskob.2022} we give examples of functions satisfying Hypothesis \ref{H2}.

We can now state the result from \cite{haskob.2022} , which established   the existence of the boundary value of
 the expansion coefficients.

\begin{theorem}[\cite{haskob.2022}] \label{thm:boundaryvelueexpcoeff} Suppose  Hypothesis \ref{H1} holds and suppose $\psi_1,\psi_2 \in L^2(\R^d)$ satisfy  Hypothesis   \ref{H2}. Then for any $E > 0$ the following limit exists    as a finite complex number
$$
 T_{n,\infty}[E \pm  i 0^+;\psi_1,\psi_2] := \lim_{\eta \downarrow 0 } T_{n,\infty}[E \pm  i \eta;\psi_1,\psi_2] .
$$
Moreover,  $z \mapsto T_{n,\infty}[z  ; \psi_1, \psi_2] $  as a function on $\C_+ := \{ z \in \C : \Im (z) > 0 \}$  has a continuous extension to a function on $\C_+ \cup  \{ z \in \C : \Re (z) > 0 \text{ and } \Im (z) = 0 \}$.
\end{theorem}
Now using  Theorem   \ref{thm:boundaryvelueexpcoeff}
to control the boundary value of the expansion coefficients  \eqref{SnL}, we obtain  the following result.

 \begin{theorem} \label{thm:duhamelmainexp2}
Suppose Hypotheses  {\bf \ref{H4}} and {\bf \ref{H1}} hold.  Let   $E  > 0 $ and $\epsilon > 0$.
Then for all
 $\psi_1,\psi_2 \in L^2(\R^3)$  satisfying Hypothesis {\bf   \ref{H2}}
 the limit
$$
\lim_{\eta \downarrow 0 }  \lim_{L \to \infty} S_{n,L}[E + i \eta; \chi ; \psi_1,\psi_2; \epsilon]    = T_n[E + i 0^+; \psi_1, \psi_2]
$$ exists, where  $S_{n,L}$ is defined in \eqref{SnL}.
\end{theorem}
The proof of Theorem \ref{thm:duhamelmainexp2} will be given in Section  \ref{sec:proofofmainthms}.

\section{Proof of the Main  Results}

  \label{sec:proofofmainthms}

In this section, we give  proofs of Theorem \ref{thm:duhamelmainexp}, Corollary \ref{thm:duhamelmainexpdens}  and Theorem   \ref{thm:duhamelmainexp3}.
We will start by outlining the basic idea of the proof of Theorem \ref{thm:duhamelmainexp}.
First we observe that, by the spectral theorem  for $\eta > 0$ we have
\begin{align} \label{expresrel00}
\inn{ \psi_1 , ( H_0 + \lambda V -  E  -  i \eta )^{-1}  \psi_2 }  & =  i  \int_0^\infty  \inn{ \psi_1 , e^{ -  i t (  H_0 + \lambda V  -E -  i \eta ) }  \psi_2 }   dt .
\end{align}
Using a so called Du Hamel expansion we will determine the time evolution appearing on the RHS of  \eqref{expresrel00}.  To control the Du Hamel expansion for large $t$, we will use the following lemma.

\begin{lemma} \label{lem:infintapprox} Let $A$ be a self-adjoint operator in a Hilbert space $\HH$
and  let $\psi_1$ and $\psi_2$ be normalized  vectors in  $\HH$.
Let $\epsilon > 0$. Then, for all $\eta > 0$ and $\tau > 0$ with  \begin{equation}\label{approxint} \tau^{-1}   \geq   \eta^{-1}  (- \ln( \eta \epsilon)),
\end{equation}
we have
\begin{equation} \label{eq:intduhammain}
\left| i  \int_0^\infty \chi_{\tau} (t) \inn{ \psi_1 , e^{ - i t (A -  E -   i \eta ) } \psi_2 }dt -  \inn{ \psi_1 , ( A -  E  -  i \eta )^{-1}  \psi_2 }
\right|   \leq  \epsilon .
\end{equation}
\end{lemma}

\begin{proof} Using \eqref{expresrel00} we estimate
\begin{align*}
& \left| i \int_0^\infty \chi_{\tau } (t) \inn{ \psi_1 , e^{ - i t (A -  E -   i \eta ) } \psi_2 }dt -  \inn{ \psi_1 , ( A -  E  -  i \eta )^{-1}  \psi_2 }
\right| \\
& \leq  \left| \int_0^\infty (\chi_{\tau } (t) - 1 ) \inn{ \psi_1 , e^{ - i t (A -  E -   i \eta ) } \psi_2 }dt \right| \\
& \leq  \left| \int_0^\infty (\chi_{\tau } (t) - 1 ) \underbrace{\| \psi_1 \|}_{\leq 1}  \underbrace{\| e^{ - i t (A-  E)} \|}_{\leq 1}   \| e^{ -  t \eta}  \| \underbrace{\| \psi_2 \|}_{\leq 1} dt \right| \\
& \leq  \left| \int_0^\infty (\chi_{\tau } (t) - 1 ) e^{- \eta t} dt
\right| \\
& \leq   \int_{\tau^{-1}}^\infty  e^{- \eta t} dt  = \frac{1}{\eta} e^{-\frac{ \eta }{\tau}  }  \leq \epsilon ,
\end{align*}
using \eqref{approxint} in the final step.
\end{proof}

The following lemma contains the expansion which we will be used for the proof of  Theorem \ref{thm:duhamelmainexp}, which is known as Du Hamel expansion. To formulate the next lemma we recall the inductive definition of the finite product  of the first $n$
operators of  a sequence of bounded operators $(A_j )_{j \in \N }$
$$
 \prod_{j=1}^{1} A_j :=A_1     ,\quad  \prod_{j=1}^{n+1} A_j :=\left(  \prod_{j=1}^{n} A_j  \right) A_{n+1}   .
$$

\begin{lemma} \label{lem:duhamel1}   Let $\lambda \in \R$, and let $D$  and $W$ be self-adjoint operators on a Hilbert space  with   $W$  bounded. Then for  $t \geq 0$
\begin{align*}
e^{ - i t (D + \lambda W)} =  \sum_{n=0}^{N-1} (- i \lambda)^n   E_n(t) + (- i \lambda)^N  F_N(t) ,
\end{align*}
where we defined
$
E_0(t)  = e^{ - i t D}
$
and for $n \geq 1$
\begin{align}
E_n(t) & :=\int_{[0,t]^n }  1_{\sum_{j=1}^n s_j \leq t}  e^{- i (t -  \sum_{j=1}^n s_j ) D }
\prod_{j=1}^n \left\{  W e^{ -i s_j D}  \right\}   d ( s_1 ,  \ldots ,  s_n ) \label{duhamel1}  \\
& =\int_{\R_+^{n}  } \int_{\R_+} \delta \left(  \sum_{j=0}^n s_j  - t \right) e^{- i s_0 D } \prod_{j=1}^n \left\{  W e^{ -i s_j D}  \right\}   ds_0   d ( s_1 ,  \ldots ,  s_n ) ,
  \label{duhamel2}
\end{align}
\begin{align} \label{Fnt}
F_n(t) :=  \int_{[0,t]^n }  1_{\sum_{j=1}^n s_j \leq t}  e^{- i (t -  \sum_{j=1}^n s_j  ) ( D + \lambda W) }  \prod_{j=1}^n \left\{  W e^{ -i s_j D}  \right\}     d ( s_1 ,  \ldots ,  s_n ).
\end{align}
where $\delta( s - a ) ds := d \delta_a(s)$ denotes the  Dirac measure at the point $a$ and the integrals in   \eqref{duhamel1}, \eqref{duhamel2}, and \eqref{Fnt} are understood as Riemann-integrals with respect to  the strong operator topology.

\end{lemma}
\begin{proof}  Let $A = D + \lambda W$.
  Differentiating $s \mapsto e^{ i s H} e^{-i s D}$, integrating from zero to $t$, and multiplying by $e^{ - i t A}$ from the right,  we find
\begin{equation} \label{eq:duhamel1}
e^{ - i t A}  = e^{- i t D}  + \int_0^t  e^{ - i (t-s) A} ( - i \lambda W ) e^{- i s D }  d s.
\end{equation}
This shows the formula for $N=1$.  The formula for general $N$ now holds inductively, since inserting   \eqref{eq:duhamel1}    into the first factor of $F_n$ implies that
$F_n = E_n + ( -i \lambda)  F_{n+1}$, which  can be seen as follows
\begin{align*}
F_n(t)
& =   \int_{[0,t]^n } 1_{\sum_{j=1}^n s_j \leq t}  e^{- i (t -  \sum_{j=1}^n s_j  ) ( D + \lambda W) }  \prod_{j=1}^n \left\{  W e^{ -i s_j D}  \right\}     d (s_1 , \cdots ,  s_n) \\
& =   \int_{[0,t]^n }  1_{\sum_{j=1}^n s_j \leq t}  e^{- i (t -  \sum_{j=1}^n s_j  )  D }  \prod_{j=1}^n \left\{  W e^{ -i s_j D}  \right\}  d (s_1 , \cdots ,  s_n)   \\
& +   \int_{[0,t]^n }  1_{\sum_{j=1}^n s_j \leq t}  \int_0^{t-\sum_{j=1}^n s_j}
 \\
& \quad \times e^{- i (t -  \sum_{j=1}^{n+1} s_j  ) ( D + \lambda W) }  (- i \lambda W ) e^{- i s_{n+1} D} \prod_{j=1}^n   \left\{  W e^{ -i s_j D}  \right\}  ds_{n+1} d (s_1 , \cdots ,  s_n)   \\
& =  E_{n}(t)     - i \lambda  \int_{[0,t]^{n+1} }
 1_{\sum_{j=1}^{n+1} s_j \leq t}   e^{- i (t -  \sum_{j=1}^{n+1} s_j  ) ( D + \lambda W) }   W  e^{- i s_{n+1} D}
 \\
 & \times \prod_{j=1}^n   \left\{  W e^{ -i s_j D}  \right\}     d (s_1 , \cdots ,  s_n , s_{n+1} )   \\
& = E_{n}(t) - i \lambda F_{n+1}(t) ,
\end{align*}
where in the last line we used a simple permutation of integration variables.
Finally observe that an  evaluation of an integral with respect to the Dirac point measure  shows that  \eqref{duhamel1}
and  \eqref{duhamel2} are equal.
\end{proof}

\begin{remark} \label{rem:logest}  { \rm  Note that  we can estimate the error term of Lemma  \ref{lem:duhamel1} by
$
| F_N(t) | \leq  |\lambda|^{N}  \| W \|^{N} \frac{t^N}{N!}  .
$
Since the  Poisson distributed random potential is  not  bounded  as a measurable function,   we need to use a different estimate and work with  expectations.
}
\end{remark}

For the proofs of the main results,  we shall make use of the  representation  in momentum space.  To this end this we shall expand with respect to the ONB $\varphi_p$ defined in \eqref{eq:deofonb}. That is, we  shall  insert the following  identity, which
holds in strong operator topology
\begin{equation} \label{fourierident}
\psi (x)= \sum_{p \in \Lambda_L^*}  \SP{\varphi_p ,  \psi}   \varphi_p  (x) =  \int_{\Lambda_L^*}  \SP{e^{  2 \pi i \inn{ p  , \cdot } } , \psi (\cdot)} e^{  2 \pi i \inn{ p , x }  } dp,  \quad x \in \Lambda_L
\end{equation}
for all $\psi \in L^2 (\Lambda_L)$, which provides  the  representation via the ONB $\{\varphi_p : p \in \Lambda_L^* \}$.

To state the next lemma, we are now going to introduce  the discrete delta function  in momentum space  for $u \in \Lambda_L^*$
\begin{equation}
\delta_*(u) = 1_{\{0\}}(u)
\end{equation}
as well as the normalized discrete delta function
\begin{equation} \label{deltaL}
\delta_{*,L}( u  ) = |\Lambda_L|  \delta_*(u),
\end{equation}
where we used the notation of the characteristic  function on a set $A$, i.e.  $1_A(x) = 1$, if $x \in A$, and $1_A(x) = 0$, if $x \notin A$.
Let $\mathcal{A}_n$ be the set of partitions of the set $ \{1,2,...,n\}$.
\begin{lemma} \label{expProd00123} For the model introduced in Section \ref{Model} we have
\begin{align}
& { \bf E}_L \left(   \prod_{j=1}^n  \hat{V}_{L,\omega}(p_j - p_{j+1} )    \right)
=  \sum_{A \in \mathcal{A}_n}  \prod_{a \in A} \left\{ m_{|a|}  \delta_{*,L} \left( \sum_{l \in a} (p_l - p_{l+1}) \right)
  \prod_{l \in a}  \hat{B}_\#(p_l - p_{l+1} ) \right\}  .  \label{eq:ofexpofrandpot2}
\end{align}
\end{lemma}
For a proof of  Lemma \ref{expProd00123}  see \cite{ErdosSalmhoferYau.2008a} or \cite{haskob.2022} respectively.

\begin{proof}[Proof of Theorem  \ref{thm:duhamelmainexp}] Let
\begin{equation} \label{eq:relaeta}
a  = \frac{  \eta}{ | \ln( \frac{\eta \epsilon}{2} )| + 1 }  .
\end{equation}
  (a)
Inserting  the Du Hamel expansion as stated in Lemma \ref{lem:duhamel1}  for $D = - \frac{\hbar^2}{2m} \Delta_L$ and $W= V_L$ into  \eqref{eq:intduhammain},  we find
\begin{align}
& \int_0^\infty \chi_a(t) \inn{ \psi_{1,\#} , e^{ - i t (D + \lambda V_L - E -  i \eta ) } \psi_{2,\#} }dt  \nonumber  \\
& =  \int_0^\infty \chi_a(t) \inn{ \psi_{1,\#} , \sum_{n=0}^{N-1} (-i \lambda)^n E_n(t) e^{ i E t - \eta t }   \psi_{2,\#} }dt
\\
&+  \int_0^\infty \chi_a(t) \inn{ \psi_{1,\#} ,  (- i \lambda)^N F_{N}(t) e^{ i E t - \eta t}   \psi_{2,\#} }dt .
\label{contapproxproof1}
\end{align}
 Let us first consider the error term, i.e., the second term in \eqref{contapproxproof1}.
Using the definition \eqref{Fnt}, interchanging integrals (which is justified by continuity in $s_j$ and the bounded domain of integration
for the $s_j$ variables), using Cauchy-Schwarz for the inner product and again for the probability measure ${\bf E}_L$  we find
  \begin{align}
& \left|  {\bf E}_L  \inn{ \psi_{1,\#} ,  F_{N}(t) e^{i E t - \eta t}   \psi_{2,\#} } \right| \nn \\
& =\left|  {\bf E}_L  \inn{ \psi_{1,\#} ,  \prod_{j=1}^N \left\{ \int_{0}^t   ds_j  \right\} 1_{\sum_{j=1}^N s_j \leq t}  e^{- i (t -  \sum_{j=1}^n s_j ) ( D + \lambda V_L) }  \prod_{j=1}^N  \left\{  V_L  e^{ -i s_j  D} \right\}
    e^{i E t - \eta t}   \psi_{2,\#} } \right| \nn  \\
    & =\left|  \prod_{j=1}^N \left\{ \int_{0}^t   ds_j  \right\}  1_{\sum_{j=1}^N s_j \leq t}  {\bf E}_L \inn{ \psi_{1,\#} ,   e^{- i (t -  \sum_{j=1}^n s_j ) ( D + \lambda V_L) }  \prod_{j=1}^N  \left\{  V_L  e^{ -i s_j  D} \right\}
    e^{i E t - \eta t}   \psi_{2,\#} } \right| \nn  \\
          & \leq  \prod_{j=1}^N \left\{ \int_{0}^t   ds_j  \right\}  1_{\sum_{j=1}^N s_j \leq t}  {\bf E}_L  \|  \psi_{1,\#} \|
            \left\|   \prod_{j=1}^N  \left\{  V_L  e^{ -i s_j  D} \right\}
    e^{i E t - \eta t}   \psi_{2,\#} \right\|   \nn  \\
      & \leq   \|  \psi_{1,\#} \|  \prod_{j=1}^N \left\{ \int_{0}^t   ds_j  \right\}  1_{\sum_{j=1}^N s_j \leq t} \left[  {\bf E}_L
            \left\|   \prod_{j=1}^N  \left\{  V_L  e^{ -i s_j  D} \right\}
    e^{i E t - \eta t}   \psi_{2,\#} \right\|^2 \right]^{1/2}    \nn  \\
 &  =  \|  \psi_{1,\#} \|    e^{ - \eta t}
 \prod_{j=1}^N \left\{ \int_{0}^t   ds_j  \right\}  1_{\sum_{j=1}^N s_j \leq t}  \Bigg[   {\bf E}_L   \left\|
     \prod_{j=1}^N  \left\{  V_L  e^{ -i s_j  D} \right\}   \psi_{2,\#}  \right\|^2  \Bigg]^{1/2} .
   \label{eq:errorerror}
\end{align}
To estimate the right hand side,  we use \eqref{fourierident} and \eqref{eq:ofexpofrandpot2}
and Fubini  (which can be justified by conditioning over the number $M$ of Poisson points and observing that the conditioned potentials are bounded
and have sufficient  decay in Fourier space, cf. Lemma
\ref{lemestfourdisc}).  Hence
\begin{align}
R&:=   {\bf E}_L   \left\|
     \prod_{j=1}^N  \left\{  V_L  e^{ -i s_j  D} \right\}   \psi_{2,\#} \right\|^2 \nn
\\
&=  \left|   {\bf E}_L   \left\langle
     \prod_{j=1}^N  \left\{  V_L  e^{ -i s_j  D} \right\}   \psi_{2,\#}  ,   \prod_{j=1}^N  \left\{  V_L  e^{ -i s_j  D} \right\}      \psi_{2,\#}  \right\rangle \right|   \nn
\\
& = \bigg| {\bf E}_L \int_{( \Lambda_L*)^{2N+1}}   \overline{\hat{\psi}_{2,\#}}(p_1)  \prod_{j=1}^N e^{  i \nu(p_j)  s_{N+1-j}} {\hat{V_L}}(p_j - p_{j+1})
 \nn
\\ & \times \prod_{j=N+1}^{2N} \hat{V_L}(p_{j } - p_{j+1}) e^{ - i \nu(p_{j+1})  s_{j-N}}
{\hat{\psi}_{2,\#}}(p_{2N+1}) )  d ( p_1 ,  \dots , p_{2N+1})   \bigg| \nn \\
&  =   \bigg|   \int_{( \Lambda_L*)^{2N+1}}  \overline{\hat{\psi}_{2,\#}(p_1) }  \hat{\psi}_{2,\#}(p_{2N+1})
e^{  i  \sum_{j=1}^N   \nu(p_j)  s_{N+1-j} } e^{  - i \sum_{j=N+1}^{2N}  \nu(p_{j+1})  s_{j-N  }}
\nn
 \\
 & \times \sum_{A \in \mathcal{A}_{2N}}  \prod_{a \in A} \left\{  m_{|a|}   \delta_{*,L} \left( \sum_{l \in a} (p_l - p_{l+1}) \right)
  \prod_{l \in a}  \hat{B}_\#(p_l - p_{l+1} )  \right\}  d ( p_1 ,  \dots , p_{2N+1})   \bigg|  \nn  \\
  &  \leq    \int_{( \Lambda_L*)^{2N+1}} \sum_{A \in \mathcal{A}_{2N}}  \bigg|    \overline{\hat{\psi}_{2,\#}(p_1) }  \hat{\psi}_{2,\#}(p_{2N+1})  \nn
 \nn \\
 & \times   \prod_{a \in A} \left\{  m_{|a|}   \delta_{*,L} \left( \sum_{l \in a} (p_l - p_{l+1}) \right)
  \prod_{l \in a}  \hat{B}_\#(p_l - p_{l+1} )  \right\}  \bigg|  d ( p_1 ,  \dots , p_{2N+1})     .
\label{boundonRRRR}
\end{align}
We introduce a change of variables given by
\begin{align} \label{changevar1}
u_0  = k_1 ,  \qquad  u_{s} = p_{s+1} - p_s,  \quad s=1,...,2N
\end{align}
to be able to resolve the discrete delta functions one at a time.
Expressing the variables $k$  in terms of the variables $u = (u_0,  \ldots u_{2N}) \in (\R^d)^{2N+1}$ we find
$$
p_j = \sum_{l=0}^{j-1} u_l .
$$
Applying this change of variables in  \eqref{boundonRRRR} we arrive at
\begin{align}
R & \leq    \int_{{\Lambda_L*}^{2N+1}}  \sum_{A \in \mathcal{A}_{2N}}   \bigg|  \overline{ \widehat{ \psi}}_{2,\#}(u_0)    \widehat{\psi}_{2,\#}  \left(\sum_{l=0}^{2N} u_l \right)   \prod_{ a \in A} \left\{   m_{|a|}  \delta_{*,L} \left( \sum_{s  \in a } u_s  \right) \prod_{s \in a}  \hat{B}_\#(-u_s) \right\}  \bigg|  \nn
\\
&\times du_0  \cdots du_{2N}   .
 \label{boundonRRRR2}
\end{align}
At this point,  we introduce second change of variables as in
\cite{haskob.2022}.
 For a given partition $A \in \mathcal{A}_{n}$ for $n=2N$ we define the set of all  indices, which are the maximum of a set $a$ by
\begin{align}
J_A := \{ \max a : a \in A\}
\end{align}
as well as its complement
\begin{align}
I_A := \{ 1 ,  \ldots ,  n \} \setminus J_A.
\end{align}
Since $A$ is a partition, for any $j \in \{1 , \ldots , n\}$ there is a unique set $a \in A$ such that $j \in a$, we denote this set by $a (j)$.
We define the map $M_A :( \R^d)^{| I_A |} \to (\R^d)^n$ as
\begin{align}\label{changevar1next}
[M_A (v)]_j :=
\begin{cases}
\qquad v_j &: j \in I_A
\\
- \sum\limits_{ l \in a (j)  \setminus \{ j \} }  v_l &: j \in J_A,
\end{cases}
\end{align}
where $v=(v_1,  \ldots , v_n)$ with $v_j \in \R^d$ for all $j \in \{1, \ldots , n \}$.
Note that \eqref{changevar1next} contains the case in which $j$ is the only element of $a(j)$ and  $[M_A (v)]_j =0$ holds.
Inserting this change of variables in  \eqref{boundonRRRR2} we find
\begin{align}
R
& \leq  \sum_{A \in \mathcal{A}_{2N}}   \int_{\Lambda_L^*} du_0 \prod_{l \in I_A} \left\{ \int_{\Lambda_L^*}  dv_l \right\}     \bigg|   \overline{\widehat{ \psi}}_{2,\#}(u_0)
   \widehat{\psi}_{2,\#}  \left( u_0 + \sum_{l=1}^{2N} [M_A(v)]_l \right) \prod_{a \in A} \{  m_{|a|}   \}
   \\
   &\times  \prod_{j  = 1}^{2N}    \hat{B}_\#(-[M_A(v)]_j)    \bigg|   \nn \\
& \leq  C^{2 N}| \mathcal{A}_{2N}|   \|  \psi_{2,\#} \|^2 ( \|  \hat{B}_\# \|_{*,\infty} + \|  \hat{B}_\# \|_{*,1} )^{2N}    ,   \label{boundonRRRR4}
\end{align}
with $C \geq 1$  such that  for all $n \in \N$  $|m_n | \leq C$, which exists by  Hypothesis \textbf{\ref{H4}}.
In the last step we used the Cauchy-Schwarz inequality for the integration over $u_0$ and
an $l^1$-bound for the integration over the $v_j$ variable for $j \in I_A$.
For compactness of notation we shall write
\begin{align} \label{eq:estonB}
 \|  \hat{B}_\# \|_{*,1,\infty} :=  \|  \hat{B}_\# \|_{*,\infty} + \|  \hat{B}_\# \|_{*,1} .
\end{align}
Obeserve that by  Lemma  \ref{lemestfourdisc}, we know that  \eqref{eq:estonB} is bounded uniformly in $L \geq 1$.
Thus, we find from \eqref{boundonRRRR4}  that
\begin{equation}
\sqrt{ R}  \leq  { \mathfrak{B}}_{2N}^{1/2}  C^{N}    \| \psi_{2,\#} \|_{2,*}    \| \hat{B}_\#  \|_{*,1,\infty}^{N}  , \label{estintermsofBell}
\end{equation}
where $\mathfrak{B}_n$ denotes the Bell number, which counts the number of partitions of a set of $n$ elements. The Bell numbers  have
the following asymptotic bound \cite{BerendTassa.2010}
\begin{equation} \label{Bellasymp00}
\mathfrak{B}_n < \left( \frac{ 0.792 n }{\ln(n+1)} \right)^n.
\end{equation}
 Now inserting   \eqref{estintermsofBell}  into   \eqref{eq:errorerror}  we find
\begin{align}
& \left|  {\bf E}_L  \inn{ \psi_{1,\#} ,  F_{N}(t) e^{- i E t - \eta t}   \psi_{2,\#} } \right|  \nn \\
&\leq \|  \psi_{1,\#} \|    e^{ - \eta t}   \nn
\int_{[0,t]^N}  1_{\sum_{j=1}^N s_j \leq t} d(s_1,  \ldots ,s_N) \nn \\
& \times    \mathfrak{B}_{2N}^{1/2}  C^{N}    \|  \psi_{2,\#} \|   \| \hat{B}_\# \|_{*,1,\infty}^{N} \nn \\
&\leq \| \psi_{1,\#} \| \|  \psi_{2,\#} \| \|  \hat{B}_\# \|_{*,1,\infty}^N  C^N \mathfrak{ B}_{2N}^{1/2} e^{- \eta t} \frac{t^N}{N!} \label{simplex},
\end{align}
where we used that the integral  in the second line coincides with the volume of the $N$-dimensional standard simplex with side length $t$, which has a volume of $\frac{t^N}{N!}$.
Using \eqref{simplex} to estimate the second term of \eqref{contapproxproof1}  yields
\begin{align*}
& \left|  \int_0^\infty {\bf E}_L  \chi_a(t) \inn{\psi_{1,\#} ,  F_{N}(t) e^{- i E t - \eta t}   \psi_{2,\#} }dt \right|   \\
 & \leq  \| \psi_{1,\#} \| \| \psi_{2,\#} \|   \left(  C \| \hat{B}_\# \|_{*,1,\infty}\right)^{N} \frac{ \mathfrak{B}_{2N}^{1/2}}{N!}   \int_0^\infty | \chi_a(t) |   t^Ne^{- \eta t }  dt    .
\end{align*}
Now  using that the support of $\chi_a$ is contained in $[- 2 /a , 2 / a ]$ we can estimate the integral and find
\begin{align}
& \left|  \int_0^\infty {\bf E}_L  \chi_a(t) \inn{\psi_{1,\#} ,  F_{N}(t) e^{- i E t - \eta t}   \psi_{2,\#} }dt \right|  \nn \\
& \leq  \| \psi_{1,\#} \| \| \psi_{2,\#} \|  \left(  C   \| \hat{B}_\# \|_{*,1,\infty}\right)^{N} \frac{ \mathfrak{B}_{2N}^{1/2}}{N!}    \int_0^{2/a}  \underbrace{| \chi_a(t) | }_{\leq 1}  t^N\underbrace{e^{- \eta t }}_{\leq 1}  dt  \nn \\
&  \leq  \| \psi_{1,\#} \| \| \psi_{2,\#} \|  \left( C  \| \hat{B}_\# \|_{*,1,\infty}  \right)^{N}  \frac{\mathfrak{B}_{2N}^{1/2}}{(N+1)!}    (2/a)^{N+1}  \nn \\
&\leq  \| \psi_{1,\#} \| \| \psi_{2,\#} \|  \left( C  \| \hat{B}_\# \|_{*,1,\infty}   \right)^{N} \left( \frac{ 1.584 N }{\ln(2N+1)} \right)^N \frac{1}{(N+1)!}    (2/a)^{N+1} \nn \\
&\leq  \| \psi_{1,\#} \| \| \psi_{2,\#} \|   \frac{2}{a} \left( \frac{3.168 C  \| \hat{B}_\# \|_{*,1,\infty}  }{a} \right)^{N} \left( \frac{  N }{\ln(2N+1)} \right)^N \frac{1}{N!}  \nn   \\
&\leq  \| \psi_{1,\#} \| \| \psi_{2,\#} \|   \frac{2}{a}  \left( \frac{3.168 C   \| \hat{B}_\# \|_{*,1,\infty}   }{a} \right)^{N} \left( \frac{  N }{\ln(2N+1)} \right)^N \frac{e^N}{\sqrt{ 2 \pi N} N ^N} \nn  \\
&\leq  \| \psi_{1,\#} \| \| \psi_{2,\#} \|   \frac{2}{a}    \left( \frac{  3.168 eC  \| \hat{B}_\# \|_{*,1,\infty} }{a\ln(2N+1)} \right)^N \frac{1}{\sqrt{ 2 \pi N} }
  \to 0    ,  \label{conterrfinal00}
\end{align}
as $N \to \infty$, where  in the third equality we made use of the asymptotics of the Bell numbers  \eqref{Bellasymp00} and in the second to last equality we applied Sterling's approximation
\begin{equation} \label{sterling00}
N! = \sqrt{ 2 \pi N} \left( \frac{N}{e} \right)^N \left( 1 + \frac{1}{12 N} + o \left(N^{-1} \right) \right) .
\end{equation}
Now, we treat the main terms,   i.e., the first  term in \eqref{contapproxproof1}. A calculation using \eqref{duhamel2}
and Fubini
shows that
\begin{align}
  & \int_0^\infty \chi_a(t) \inn{ \psi_{1,\#} ,  E_n(t) e^{- i E t - \eta t }   \psi_{2,\#} }dt  \label{contmainfinal00}  \\
& =
 \int_0^\infty \chi_a(t)   \int_{\R_+^{n+1} } \delta \left( t - \sum_{j=0}^n s_j \right)  \nn \\
 & \times \inn{ \psi_{1,\#} ,  e^{- i s_0 (D - E - i \eta) } \prod_{j=1}^n  \left\{  V_L  e^{ -i s_j  (D-E-i \eta)} \right\} \psi_{2,\#} } d ( s_0,  s_1 \cdots s_n) d t \nn \\
& =
 \int_{\R_+^{n+1} } \chi_a \left(   \sum_{j=0}^n s_j   \right)
 \inn{ \psi_{1,\#} ,  e^{- i s_0 (D - E - i \eta) } \prod_{j=1}^n  \left\{   V_L  e^{ -i s_j  (D-E-i \eta)} \right\} \psi_{2,\#} } d ( s_0,  s_1 \cdots s_n)  \nn \\
& =   \int_\R     \widehat{ \chi}_a(\alpha)  \int_{\R_+^{n+1} }   e^{   2 \pi  i  \sum_{j=0}^n s_j  \alpha  } \nn
\\
 & \times \inn{ \psi_{1,\#} ,e^{- i s_0 (D -  E  - i \eta)}  \prod_{j=1}^n  \left\{  V_L  e^{ -i s_j  (D-E-i \eta)} \right\} \psi_{2,\#}} d ( s_0,  s_1 \cdots s_n) d \alpha \nn \\
& =  (-i )^{n+1}    \int_\R    \widehat{ \chi}_a(\alpha)    \nn
 \\ &\times \inn{ \psi_{1,\#}  ,  (D   -  E - 2 \pi \alpha - i \eta )^{-1}    \prod_{j=1}^n  \left\{  V_L  (D - E -  2 \pi \alpha -  i \eta  )^{-1}  \right\}  \psi_{2,\#} } d \alpha   ,  \nn
\end{align}
where in the second to last  step we used
$$
\int_0^\infty e^{ - i s x - \eta s } ds = \frac{ 1 }{ i   x  +    \eta } = \frac{ -i }{   x  - i   \eta } , \quad \eta > 0 .
$$
Calculating the expectation of \eqref{contmainfinal00} using    \eqref{defoffinT} we find
\begin{align}
& {\bf E}_L   \int_0^\infty \chi_a(t) \inn{ \psi_{1,\#} ,  E_n(t) e^{- i E t - \eta t }   \psi_{2,\#} }dt
 =   (-i)^{n+1}   \int_\R    \widehat{ \chi}_a(\alpha)  T_{n,L}[ E + i \eta +  2 \pi \alpha; \psi_1, \psi_2 ] .   \label{contmainfinal000}
\end{align}

Now  we collect the previous estimates.  Using  first  the  definition of $S_{n,L}$ given in \eqref{SnL}  together with   Eq. \eqref{contmainfinal000}, second  Eq. \eqref{contapproxproof1}
and  third  Lemma  \ref{lem:infintapprox}  for  $\tau = a$ and  \eqref{conterrfinal00}, we find
\begin{align}
& \left| {\bf E}_L \inn{ \psi_{1,\#} , (H_\lambda - E -  i \eta )^{-1}      \psi_{2,\#} }  -    \sum_{n=0}^{N-1}  \lambda^n S_{n,L}[ E + i   \eta; \chi; \psi_1, \psi_2 ; \epsilon ]    \right| \nn  \\
& = \Bigg| {\bf E}_L \inn{ \psi_{1,\#} , (H_\lambda - E -  i \eta )^{-1}      \psi_{2,\#} }  \nn
\\
&-  i  \sum_{n=0}^{N-1} (-i  \lambda)^n  {\bf E}_L  \int_0^\infty \chi_a(t) \inn{ \psi_{1,\#} ,  E_n(t) e^{- i E t - \eta t }   \psi_{2,\#} }dt    \Bigg| \nn  \\
& \leq
 \left| {\bf E}_L \inn{ \psi_{1,\#} , (H_\lambda - E -  i \eta )^{-1}      \psi_{2,\#} }  -
 i  {\bf E}_L     \int_0^\infty \chi_a(t) \inn{ \psi_{1,\#} , e^{ - i t (D + \lambda V_L - E -  i \eta ) } \psi_{2,\#} }dt   \right| \nn \\
& +  \left|   \lambda^N  {\bf E}_L  \int_0^\infty \chi_a(t) \inn{ \psi_{1,\#} ,  F_N(t) e^{- i E t - \eta t }   \psi_{2,\#} }dt    \right|  \nn \\
& \leq  \frac{\epsilon}{2}   \| \psi_{1,\#}\|  \| \psi_{2,\#} \| +
  \| \psi_{1,\#} \| \| \psi_{2,\#} \|  \frac{2}{a} \left( \frac{  3.168 eC \| \hat{B}_\# \|_{*,1,\infty}  }{a\ln(2N+1)} \right)^N \frac{1}{\sqrt{ 2 \pi N} }  ,
\end{align}
as oberved after  \eqref{conterrfinal00}  the second term on the right hand side tends to zero for large $N$. This shows   \eqref{firsteqmaincont007}.
 \end{proof}

We are now going to prove Corollary \ref{thm:duhamelmainexpdens}.

\begin{proof}[Proof of   Corollary \ref{thm:duhamelmainexpdens}]
Let $\eta > 0$,  $E > 0$ and $\epsilon > 0$ and $\lambda_0 > 0$.
We are going to show below that there exists $\kappa > 0$ such that
\begin{align} \label{eq:basicErrorkappa}
& \left| \frac{1}{|\Lambda_L|} {\bf E}_L  {\rm Tr}_L f_{E,\eta} ( H_\lambda)  - \int_{|q| \leq \kappa,*} {\bf E}_L  \inn{ \varphi_q,  f_{E,\eta} ( H_\lambda)  \varphi_q} dq \right|  \nonumber \\
& =  \left| \int_{|q| >  \kappa,*} {\bf E}_L  \inn{ \varphi_q,  f_{E,\eta} ( H_\lambda)  \varphi_q} dq  \right|  \leq \frac{\epsilon}{2}.
\end{align}
With this the statement, i.e. , Eq. \eqref{firsteqmaincont007D},  is concluded using Theorem \ref{thm:duhamelmainexp},  as follows:
Explicitly, let $M_\kappa := \sup_{L \geq 1} \int_{|q| \leq \kappa, *} 1 dq$. Clearly $M_\kappa < \infty$.
 By   Theorem \ref{thm:duhamelmainexp} for $\epsilon' = \frac{\epsilon}{ 2 M_\kappa} >0$ we can choose an $N = N( \epsilon',\eta,\lambda_0)$ such that
\begin{align}  \label{firsteqmaincont007Ap}
& \left| {\bf E}_L \inn{ \psi_{1,\#} , (H_\lambda - E -  i \eta )^{-1}      \psi_{2,\#} }  -   \sum_{n=0}^{N}  \lambda^n S_{n,L}[ E + i   \eta; \chi; \psi_1, \psi_2 ; \epsilon ]    \right| \nonumber \\
&  \leq \frac{\epsilon}{ 2 M_\kappa}  \| \psi_{1,\#} \|  \| \psi_{2,\#} \|  .
\end{align}
Now using  $ f_{E,\eta}(H_\lambda)  = {\rm Im} \frac{1}{ H_\lambda - E  + i \eta}  $ it follows using \eqref{eq:basicErrorkappa},  \eqref{firsteqmaincont007Ap},  and  the triangle inequality  that
\begin{align*}
& \text{LHS of }   \eqref{firsteqmaincont007D}\\
 &  \leq  \left| \frac{1}{|\Lambda_L|} {\bf E}_L  {\rm Tr}_L f_{E,\eta} ( H_\lambda)  - \int_{|q| \leq \kappa,*} {\bf E}_L  \inn{ \varphi_q,  f_{E,\eta} ( H_\lambda)  \varphi_q} dq \right|  \\
& +   \left|  \int_{|q| \leq \kappa,*} {\bf E}_L  \inn{ \varphi_q,  f_{E,\eta} ( H_\lambda)  \varphi_q} dq      -
 \int_{|q| \leq \kappa,*} \sum_{n=0}^{N} \lambda^n  {\rm Im} S_{n,L}[ E + i   \eta; \chi; \varphi_{q}, \varphi_q ; \epsilon ] dq
 \right|   \\
& \leq \frac{\epsilon}{2} +   \bigg| {\rm Im} \bigg[  \int_{|q| \leq \kappa,*} {\bf E}_L  \inn{ \varphi_q, ( H_\lambda- E - i \eta)^{-1}  \varphi_q} dq
\\
&-  \int_{|q| \leq \kappa,*} \sum_{n=0}^{N} \lambda^n   S_{n,L}[ E + i   \eta; \chi; \varphi_{q}, \varphi_q ; \epsilon ] dq  \bigg]
 \bigg| \\
 &  \leq \frac{\epsilon}{2}  +   \int_{|q| \leq \kappa,*} 1 dq \frac{\epsilon      }{ 2 M_\kappa }= \epsilon .
\end{align*}
Thus, it remains
to show \eqref{eq:basicErrorkappa}. To this end,   let
 $D_L =  - \frac{\hbar^2}{2m} \Delta_L$.
 First observe that by the support assumption of $v_\alpha$ we have for $\lambda \geq 0$ that
with  $D_L + \lambda V \geq 0$.
By the square inequality, we have for all $x \in \R$ and $\eta > 0$
\begin{align*}
(x+1)^2 & = (x-E)^2 + 2 (x-E)(1+E) + (1+E)^2 \leq 2 (x-E)^2 + 2 (1+E)^2 \\
& \leq  2 \left( (x-E)^2 +\eta^2  \right) +  2 (1+E)^2 \frac{\eta^2}{\eta^2} + \frac{2(x-E)^2 (1+E)^2}{\eta^2}
\\ &= 2\left(  (x-E)^2 +\eta^2 \right) \left( 1 +  \frac{(1+E)^2}{\eta^2} \right)  .
\end{align*}
Hence,  by shifting,  multiplying by $\eta$ and with $C_{\eta,E} = 2 \left(\eta  + \frac{(1+E)^2 }{ \eta} \right) $, we find
\begin{align*}
f_{E,\eta}(x) = \frac{\eta }{ (x-E)^2 +\eta^2 }  \leq \frac{ C_{\eta,E} }{(x+1)^{2}}
\end{align*}
Thus, we find using the spectral theorem
\begin{align} \label{eq:FEetaest}
 f_{E,\eta}(H_\lambda) \leq \frac{ C_{\eta,E} }{(D_L + \lambda V + 1)^{2}} .
\end{align}
Next, we estimate the trace  of the right hand side of \eqref{eq:FEetaest}, when projected onto high momenta.
Now \eqref{eq:FEetaest}, positivity and writing
\muun{
\label{Ikl}
 \mathcal{I}_{\kappa,L} := \frac{1}{|\Lambda_L|} {\bf E}_L  \sum_{|p| \geq \kappa} \SP{ \varphi_p , \frac{1}{ (D_L + \lambda V  + 1 )^2} \varphi_p }
}
yield
\muun{
\label{CIkl}
\left| \int_{|q| >  \kappa,*} {\bf E}_L  \inn{ \varphi_q,  f_{E,\eta} ( H_\lambda)  \varphi_q} dq  \right|  \leq C_{\eta , E} \mathcal{I}_{\kappa,L}.
}
Applying the second resolvent identity  to \eqref{Ikl}, we find
\begin{align*}
 \mathcal{I}_{\kappa,L}
 &= \frac{1}{|\Lambda_L|}  {\bf E}_L  \sum_{|p| \geq \kappa} \SP{ \varphi_p ,\frac{1}{ D_L  +1} \frac{1}{ D_L + \lambda V  + 1 } \varphi_p } \\
& +  \frac{1}{|\Lambda_L|} {\bf E}_L  \sum_{|p| \geq \kappa} \SP{ \varphi_p ,\frac{1}{ D_L  + 1 }  \lambda V   \frac{1}{ (D_L + \lambda V + 1 )^2} \varphi_p } \\
& =: \mathcal{E}_{{\rm I}}+ \mathcal{E}_{{\rm II}} .
\end{align*}
Using the Cauchy Schwarz inequality applied to the trace in the first step and in the second that by the positivity assumption $(D_L + \lambda V + 1 )^{-1} \leq 1 $ holds   we find
\begin{align*}
  |  \mathcal{E}_{{\rm II}} |  & =  \left| \frac{1}{|\Lambda_L|}  {\bf E}_L  \sum_{|p| \geq \kappa} \SP{ \varphi_p ,\frac{1}{ D_L  +1}  \lambda V  \frac{1}{ (D_L + \lambda V + 1)^2} \varphi_p } \right|  \\
& \leq \frac{1}{|\Lambda_L|} \left(  \sum_{|p| \geq \kappa} \SP{ \varphi_p ,\frac{1}{ D_L  +1} { \bf E}_L  (\lambda V)^2 \frac{1}{ D_L  +1}   \varphi_p }\right)^{1/2} \\
& \times \left( { \bf E}_L  \sum_{|p| \geq \kappa} \SP{ \varphi_p , \frac{1}{ (D_L + \lambda V  +1 )^4}    \varphi_p } \right)^{1/2}  \\
& \leq \frac{1}{|\Lambda_L|} \left(  \sum_{|p| \geq \kappa} \SP{ \varphi_p ,\frac{1}{ D_L  +1 } { \bf E}_L  (\lambda V)^2 \frac{1}{ D_L  +1 }   \varphi_p } \right)^{1/2} \\
& \times\left( { \bf E}_L  \sum_{|p| \geq \kappa} \SP{ \varphi_p , \frac{1}{ (D_L + \lambda V   + 1)^2}    \varphi_p } \right)^{1/2}.
\end{align*}
To show that ${ \bf E}_L  (\lambda V)^2$ is uniformly bounded    in $L \geq 1$, we have
\begin{align}\label{Vsquarecalc}  {\bf E}_L     V^2   &  =  {\bf E}_L  \left(  \sum_{\gamma=1}^M    v_\gamma  B_\#(\cdot - y_{L,\gamma}) \right)^2 \nn \\
& =   {\bf E}_L  \left(  \sum_{\gamma=1}^M \sum_{\beta=1}^M   v_\gamma v_\beta  B_\#(\cdot - y_{L,\gamma})
B_\#(\cdot - y_{L,\beta}) \right)  \nn \\
& =   {\bf E}_M \left(  \sum_{\gamma, \beta \in \{1,....,M\} :  \gamma  \neq \beta }   {\bf E}_v  (v_\gamma v_\beta)  {\bf E}_y B_\#(\cdot - y_{L,\gamma})
B_\#(\cdot - y_{L,\beta}) \right)  \nn \\
& +     {\bf E}_M  \left(  \sum_{\gamma =1  }^M   {\bf E}_v  (v_\gamma^2)  {\bf E}_y B_\#(\cdot - y_{L,\gamma})^2.
 \right)
\end{align}
To calculate  the first term on the right hand side, we use Fubinis theorem and \eqref{esy:3.240} with $\gamma \neq \beta$
\begin{align}  \label{infsquarVest2}
& {\bf E}_y \left( B_\#(x - y_{L,\gamma})
B_\#(x - y_{L,\beta})  \right) \nonumber  \\
& =  {\bf E}_y  \int_{\Lambda_L^*} \int_{\Lambda_L^*} \hat{B}_\#(k) \hat{B}_\#(k') e^{2 \pi  i k \cdot (x - y_{L,\gamma})}
  e^{2 \pi i k' \cdot (x - y_{L,\beta})} dk dk' \nonumber  \\
&=
\frac{1}{|\Lambda_L|^2}  \int_{\Lambda_L^*} \int_{\Lambda_L^*} \hat{B}_\#(k) \hat{B}_\#(k') \int_{(\Lambda_L)^2}  e^{2 \pi  i k \cdot (x - y_1)}  e^{2 \pi i k' \cdot (x - y_2)} dy_1 dy_2  dk dk' \nn \\
& =   \int_{\Lambda_L^*} \int_{\Lambda_L^*} \hat{B}_\#(k) \hat{B}_\#(k') \delta_{k,0} \delta_{k',0}  dk dk'  \nonumber  \\
&  =   \frac{1}{|\Lambda_L|^2} \hat{B}_\#(0) \hat{B}_\#(0)  .
\end{align}
and for the second sum
\begin{align} \label{infsquarVest1}
& {\bf E}_y \left( B_\#(x - y_{L,\gamma})
B_\#(x - y_{L,\gamma})  \right) \nonumber  \\
& =  {\bf E}_y  \int_{\Lambda_L^*} \int_{\Lambda_L^*} \hat{B}_\#(k) \hat{B}_\#(k') e^{2 \pi  i k \cdot (x - y_{L,\gamma})}
  e^{2 \pi i k' \cdot (x - y_{L,\gamma})} dk dk'  \nonumber  \\
& =  \int_{\Lambda_L^*} \int_{\Lambda_L^*} \hat{B}_\#(k) \hat{B}_\#(k') e^{2 \pi  i k \cdot x }
  e^{2 \pi i k' \cdot x } \delta_{0,k+k'} dk dk'  \nonumber  \\
& =   \frac{1}{|\Lambda_L|} \int_{\Lambda_L^*} \hat{B}_\#(k) \hat{B}_\#(-k)   dk \nonumber  \\
& =   \frac{1}{|\Lambda_L|} \int_{\Lambda_L} | B_\#(x)|^2 dx  .
\end{align}
Now inserting \eqref{infsquarVest1}   and \eqref{infsquarVest2}  into
\eqref{Vsquarecalc},   for $\beta \neq \gamma$ and using that  $\{v_\gamma \}_{ \gamma \in \N}$ are i.i.d.  and that by Lemma \ref{lemP}, ${\bf E}_M M = |\Lambda_L|$ and  ${\bf E}_M (M(M-1)) = |\Lambda_L|^2$ hold, we  find
\begin{align}\label{Vsquarecalc2}
 {\bf E}_L     V^2   &
\leq {\bf E}_M (M(M-1)) {\bf E}_v (v_\gamma v_\beta) |\Lambda_L|^{-2}  \hat{B}_\#(0) \hat{B}_\#(0)  \nonumber
\\ & + {\bf E}_M M  {\bf E}_v (v_\gamma^2  )|\Lambda_L|^{-1} \int_{\Lambda_L} | B_\#(x)|^2 dx  \nonumber  \\
& \leq  ({\bf E}_v v_\gamma)^2  \hat{B}_\#(0) \hat{B}_\#(0)  + {\bf E}_v (v_\gamma^2  ) \int_{\Lambda_L} | B_\#(x)|^2 dx  .
\end{align}
Thus let
\begin{align} \label{Rkl2}
\mathcal{R}_{\kappa,L} :=  \frac{1}{|\Lambda_L|} \sum_{p \in \Lambda_L^* :  |p| \geq \kappa} \SP{ \varphi_p ,\frac{1}{ (D_L  +1)^2}    \varphi_p }
= \int_{|p| \geq \kappa,*} \frac{1}{ ( \frac{1}{2} p^2  + 1 )^2 }
 .
\end{align}
Now  using  the Fourier transfrom an elementary estimate
 shows  that there exists a constant  $c$ such that for all   $L \geq 1$ and $\kappa \geq 1$ we can estimate RHS of \eqref{Rkl2} in terms of an improper Riemann-integral, cf. Lemma  \ref{lem:discrriem},  by
 \begin{align*}
\mathcal{R}_{\kappa,L} = \int_{|p| \geq \kappa,*} \frac{1}{ ( \frac{1}{2} p^2  + 1 )^2 }  dp  \leq
   \int_{|p| \geq \kappa} \ \frac{c}{ ( \frac{1}{2} p^2  + 1 )^2 }  dp.
\end{align*}
Note that the  right hand side is finite for $d \leq 3$.
Hence,  by dominated convergence
\begin{align} \label{convF} \mathcal{R}_{\kappa,L} \to 0 \end{align}
 as $\kappa \to \infty$ uniformly in $L \geq 1$.
Thus we find using Cauchy-Schwarz and the above bounds
\begin{align*}
&  \mathcal{I}_{\kappa,L}  \leq \mathcal{I}_{\kappa,L}^{1/2} \mathcal{R}_{\kappa,L}^{1/2} +  [{ \bf E}_L  (\lambda V)^2 ]^{1/2}
  \mathcal{I}_{\kappa,L}^{1/2}  \mathcal{R}_{\kappa,L}^{1/2}
\end{align*}
Dividing by $\mathcal{I}_{\kappa,L}^{1/2}$ and squaring shows that
\begin{align} \label{Vsquarecalc44}
&  \mathcal{I}_{\kappa,L} \leq  \left(1   +  [{ \bf E}_L  (\lambda V)^2 ]^{1/2}
 \right)^2  \mathcal{R}_{\kappa,L} .
\end{align}
Now the right hand side tends to zero uniformly in $L \geq 1 $ by  \eqref{convF} and  \eqref{Vsquarecalc2} and therefore
\muun{
\label{Ikl0}
 \mathcal{I}_{\kappa,L}  \to 0
}
as $\kappa \to 0$.
Finally \eqref{eq:basicErrorkappa}  follows from inserting \eqref{Ikl0} into \eqref{CIkl}.
\end{proof}

 Next we are ready to give the proof of the  regularity of the expansion coefficients.

  \begin{proof}[Proof of Theorem   \ref{thm:duhamelmainexp3}]
From \eqref{inflimS} we find
\begin{align}
  &\lim_{L \to \infty} S_{n,L}[E + i \eta; \chi ; \psi_1,\psi_2;\epsilon] \nn  \\
  &  =  \int_\R  \widehat{ \chi}_a(\alpha)  T_{n,\infty}[E +2 \pi \alpha +  i \eta ;  \psi_1 , \psi_2 ] d \alpha    \nn   \\
     &  =  \int_\R   1_{ |  E + 2 \pi \alpha  | \leq  E/2   }   \widehat{ \chi}_a(\alpha)  T_{n,\infty}[E +2 \pi \alpha +  i \eta ;  \psi_1 , \psi_2 ] d \alpha  \nn
     \\&+   \int_{2 \pi \alpha \leq  - \frac{3 E }{2} }     \widehat{ \chi}_a(\alpha)  T_{n,\infty}[E +2 \pi \alpha +  i \eta ;  \psi_1 , \psi_2 ] d \alpha \nn
       \\&+   \int_{- \frac{E }{2 } < 2 \pi \alpha  }      \widehat{ \chi}_a(\alpha)  T_{n,\infty}[E +2 \pi \alpha +  i \eta ;  \psi_1 , \psi_2 ] d \alpha  ,  \label{convergenceofS}
\end{align}
where the second equality follows from a simple partition of $\R$ into three intervals.
The last term   in  \eqref{convergenceofS}   has the following limit as $\eta \downarrow 0$ recalling the relation in Eq.  \eqref{eq:relaeta}
\begin{align}
 & \lim_{\eta \downarrow 0}  \int_{- \frac{E }{2 (2\pi)} < \alpha  }      \widehat{ \chi}_a(\alpha)  T_{n,\infty}[E +2 \pi \alpha +  i \eta ;  \psi_1 , \psi_2 ] d \alpha \nn  \\
   & =   \lim_{\eta \downarrow 0}
  \int_\R      \widehat{ \chi}(\alpha)  1_{ - \frac{E}{2(2\pi)} < a \alpha}  T_{n,\infty}[E +2 \pi a  \alpha +  i \eta ;  \psi_1 , \psi_2 ] d \alpha \nn \\
  & =  T_{n,\infty}[E  +  i 0 ;  \psi_1 , \psi_2 ]  , \label{duhamelest1}
\end{align}
where the first equality follows by a simple change of variables and for  the second equality we used Theorem \ref{thm:boundaryvelueexpcoeff}, dominated convergence and $\int_\R \widehat{\chi}(\alpha) d \alpha = \chi(0)=1$. To estimate the first term of the right hand side of     \eqref{convergenceofS}  we use the   bound
\begin{equation} \label{eq:estfprS1} |T_{n, \infty}[\cdot + i \eta ; \psi_1 , \psi_2  ] | \leq  C_n \| \psi_{1} \| \| \psi_{2} \|  |\eta|^{-n-1} ,  \end{equation}
which holds by  Theorem \ref{thm:boundaryvelueexpcoeff}.
Since $\chi$ is smooth with compact support its Fourier transform decays faster than any polynomial.
More explicitly,  for any $m \in \N$ there exists a constant  $\tilde{c}_m$ such that
\begin{align}
| \hat{\chi}_a(s) |   & = \left|    \int_\R  e^{-  2 \pi i  s t } \chi(at) dt \right|  = \left|      \frac{1}{a} \int_\R  e^{- 2\pi i (s/a) t } \chi(t) dt \right| \nn  \\
&  \leq \frac{1}{a}  \frac{\tilde{c}_m}{1+ |s/a|^m } =    \frac{\tilde{c}_m a^{m-1}}{a^m+ |s|^n }  \leq \frac{\tilde{c}_m a^{m-1}}{ |s|^m }  . \label{eq:decayofchi}
\end{align}
We find   using  \eqref{eq:estfprS1} and  \eqref{eq:decayofchi} with $m=n+3$ and  \eqref{eq:relaeta}    that
\begin{align}
 & \bigg| \int_\R    1_{ |  E + 2 \pi \alpha  | \leq  E/2   }   \widehat{ \chi}_a(\alpha)  T_{n,\infty}[E +2 \pi \alpha +  i \eta ;  \psi_1 , \psi_2 ] d \alpha \bigg| \nn \\
 & \leq C_n \| \psi_{1} \| \| \psi_{2} \| | \eta|^{-n-1}  \int_\R    1_{ |  E + 2 \pi \alpha  | \leq  E/2   }  | \widehat{ \chi}_a(\alpha) | d \alpha  \nn \\
 & \leq C_n \| \psi_{1} \| \| \psi_{2} \| | \eta|^{-n-1}  \int_{E/(4\pi)}^{3E/(4 \pi)}
\frac{\tilde{c}_m a^{m-1}}{ |\alpha|^m } d \alpha  \nn \\
 & \leq  C_n  \| \psi_{1} \|  \| \psi_{2} \|  | \eta|^{-n-1}   \frac{ E }{2\pi}  \frac{\tilde{c}_m a^{m-1}}{( E / 4 \pi )^m } \nn  \\
 & \leq \frac{(4 \pi)^{n+3}C_n\tilde{c}_m}{2 \pi }  \| \psi_{1} \|  \| \psi_{2} \|      E^{-n-2}   \frac{|\eta| }{( |\ln(\epsilon \eta/2)|+1)^{n+3}}  \to 0
\label{duhamelest2}
\end{align}
as $\eta \downarrow 0$. To estimate the second  term  in  \eqref{convergenceofS}  we use for $s > 0$   the   bound
$|T_{n,\infty}[ - s  ; \psi_1 , \psi_2  ] | \leq  C_n \| \psi_{1} \| \| \psi_{2} \|  s^{-n} $ from Theorem \ref{thm:boundaryvelueexpcoeff}.
This gives
\begin{align}
& \left| \int_{2 \pi \alpha \leq  - \frac{3 E }{2} }     \widehat{ \chi}_a(\alpha)  T_{n,\infty}[E +2 \pi \alpha +  i \eta ;  \psi_{1} , \psi_{2} ] d \alpha \right|  \nn \\
 & \leq C_n \| \psi_{1} \| \| \psi_{2} \| (E/2)^{-n} \int_{2 \pi \alpha \leq  - \frac{3 E }{2} }    | \widehat{ \chi}_a(\alpha) d \alpha | \nn \\
 & \leq C_n \| \psi_{1} \| \| \psi_{2} \| (E/2)^{-n} \int_{2 \pi \alpha \leq  - \frac{3 E }{2a} }    | \widehat{ \chi}(\alpha) | d \alpha \to 0 ,
\label{duhamelest3}
\end{align}
where limit as  $ a \downarrow 0$  (and hence as $\eta \downarrow 0$) follows from dominated convergence.
Now the claim follows by inserting \eqref{duhamelest1},  \eqref{duhamelest2},  \eqref{duhamelest3} into
 \eqref{convergenceofS}.
\end{proof}

\appendix

\section{ Appendix}
\label{PPL}

The following lemma is used in the proof of Corollary \ref{thm:duhamelmainexpdens}.
\begin{lemma}[\cite{haskob.2022}]\label{lemP}  For a Poisson distributed random variable  $N$ with mean $\lambda$ we
 have for any $k \in \N$
\begin{align}
\E \left[ \prod_{j=0}^{k-1} (N-j) \right]
& = \lambda^k . \label{eq:poissonexp}
\end{align}
\end{lemma}


\label{sec:prooasymexp}

The following Lemma will be used for the proof of Theorem  \ref{thm:duhamelmainexp}.
For this we introduce  the notation  
 \begin{align}
 \label{Klammern}
 \langle x \rangle = (1 + x^2)^{1/2}.
 \end{align} 
\begin{lemma}[\cite{haskob.2022}] \label{lemestfourdisc}    There exists a constant $C_d$ (explicitly given in the proof)
 and  a constant $\tilde{C}_d$ such that for all $L \geq 1$ and  $f \in \mathcal{S}(\R^d)$ which are 
symmetric  with respect to the reflections $\rho_j : (x_1,...,x_j,...,x_d) \mapsto (x_1,...,-x_j,...,x_d)$ for all $j=1,...,d$ (i.e.  $f \circ \rho_j = f$ for all $j=1,...,d$), 
we have 
\begin{align}  \label{eq:dgenallk0} 
\langle  k_1 \rangle^2 \cdots  \langle k_d \rangle^2 | \widehat{f}_\#(k) |  \leq C_d  \sum_{\substack{\alpha \in \N_0^d : \\  \alpha_j \leq 2}} \sup_{x \in \R^d}  |  \langle x \rangle^{2d} \partial^\alpha f(x) | 
 \end{align} 
for all $k=(k_1, \ldots , k_d) \in \R^d$.
Furthermore $\widehat{f}_\# \in \ell_*^1(\Lambda_L^*)$,  and 
 \begin{align} \label{secondRest0} 
\| \widehat{f}_\# \|_{*,1} \leq \tilde{C}_d \sum_{\substack{ \alpha \in \N_0^d :  \\ \alpha_j \leq 2}} \sup_{x \in \R^d}| \langle x \rangle^{2d} \partial^\alpha f(x) |  . \end{align} 
\end{lemma}

We now present two estimates, while the second one is an estimate on a discrete integrals, i.e., infinite sums.

Let us first mention a basic inequality. For any $z \in \C$ we have the elementary bound
\begin{equation} \label{ln-1}
| z^{-1}| \leq \left| {\rm Im} z \right|^{-1} .
\end{equation}

The following lemma is used in the proof of  Corollary \ref{thm:duhamelmainexpdens}.

\begin{lemma}[\cite{haskob.2022}]
\label{lem:discrriem}  Let $I  = [-c,c]^d$ and $I_j$, $j=1,...,N$, be   a partition (up to boundaries) of $I$  into translates of a square  $Q$ which is centered at the
origin. Let $\xi_j  \in  I_j$
and $ \Delta x_j = | I_j|$. Suppose $g \geq 0$ is a Riemann integrable function on $I$  and we have
\begin{equation} \label{eq:estonint}
\sup_{ \xi \in I_j} f(\xi) \leq \inf_{ \xi \in I_j} g(\xi)  .
\end{equation}
Then
\begin{align*}
  \left|  \sum_{i=1}^N f(\xi_j) \Delta x_j  \right| \leq  \int_I g(x) dx  .
\end{align*}
\end{lemma}
\begin{proof}  The statement follows directly from the theory of Riemann integration.
\end{proof}

\printbibliography

\end{document}